\documentclass{PoS}

\usepackage{graphicx}
\usepackage{amsmath}
\usepackage{amssymb}
\newcommand{\be}{\begin{equation}}
\newcommand{\ee}{\end{equation}}
\newcommand{\bea}{\begin{eqnarray}}
\newcommand{\eea}{\end{eqnarray}}
\newcommand{\bi}{\begin{itemize}}
\newcommand{\ei}{\end{itemize}}
\newcommand{\ben}{\begin{enumerate}}
\newcommand{\een}{\end{enumerate}}
\newcommand{\bt}{\begin{tabbing}}
\newcommand{\et}{\end{tabbing}}
\newcommand{\nn}{\nonumber}

\newcommand{\pp}{{p^\prime}}
\newcommand{\bfp}{{\bf p}}
\newcommand{\bfpp}{{{\bf p}^\prime}}

\newcommand{\bfx}{{\bf x}}
\newcommand{\bfxp}{{{\bf x}^\prime}}

\newcommand{\bfz}{{\bf 0}}
\newcommand{\dt}{{\Delta t}}
\newcommand{\dtp}{{\Delta t^\prime}}
\newcommand{\tsrc}{{t_{\rm src}}}
\newcommand{\bfxsrc}{{{\bf x}_{\rm src}}}
\newcommand{\vp}{v^\prime}

\newcommand{\bfppp}{{\bf p}_\perp^\prime}
\newcommand{\bfppnp}{{\bf p}_{\not\perp}^\prime}

\title{
   \begin{picture}(0,0)(0,0)%
   \put(350,75){\makebox(0,0)[l]{\textnormal{\normalsize KEK-CP-369}}}%
   \put(350,62){\makebox(0,0)[l]{\textnormal{\normalsize OU-HET-983}}}%
   \end{picture}
   $B\!\to\!D^{(*)}\ell\nu$ form factors from $N_f\!=\!2+1$ QCD
   with M\"obius domain-wall quarks
}

\ShortTitle{
   $B\!\to\!D^{(*)}\ell\nu$ form factors from $N_f\!=\!2+1$ QCD
   with M\"obius domain-wall quarks
}

\author{
   JLQCD Collaboration:
   \speaker{T.~Kaneko}$^{a,b}$\thanks{E-mail: takashi.kaneko@kek.jp},
   Y.~Aoki$^a$,
   B.~Colquhoun$^a$,
   H.~Fukaya$^c$,
   S.~Hashimoto$^{a,b}$
   \\
   \\
   \\
   \llap{$^a$}
   High Energy Accelerator Research Organization (KEK),
   Ibaraki 305-0801, Japan 
   \\
   \llap{$^b$}
   School of High Energy Accelerator Science,
   SOKENDAI (The Graduate University for Advanced Studies),
   Ibaraki 305-0801, Japan
   \\
   \llap{$^c$}
   Department of Physics, Osaka University, 
   Osaka 560-0043, Japan
}

\abstract{
We report on our study of the $B \to D^{(*)}\ell\nu$ semileptonic decays
at zero and nonzero recoils in 2+1 flavor QCD.
The M\"obius domain-wall action is employed for light,
charm and bottom quarks at lattice cutoffs $a^{-1} = 2.5$ and 3.6 GeV.
We take bottom quark masses up to $\approx$\,2.4 times the physical charm mass
to control discretization effects.
The pion mass is as low as $M_\pi\!\sim\!310$~MeV. 
We present our preliminary results for the relevant form factors
and discuss the violation of heavy quark symmetry, 
which is a recent important isuue on the long-standing tension
in the Cabibbo-Kobayashi-Maskawa matrix element $|V_{cb}|$
between the exclusive and inclusive decays.
}

\FullConference{The 36th Annual International Symposium
                on Lattice Field Theory - LATTICE2018\\
		22-28 July, 2018\\
		Michigan State University, East Lansing, Michigan, USA.}

\begin{document}

%// introduction ===============================================================

\section{Introduction}

The $B \to D^{(*)}\ell\nu$ semileptonic decays provide
a determination of the Cabibbo-Kobayashi-Maskawa (CKM) matrix element
$|V_{cb}|$.
There has, however, been a long-standing tension
with $|V_{cb}|$ from the inclusive decay~\cite{HFLAV},
which has to be resolved towards
an unambiguous interpretation of precise experimental data
from forthcoming experiments at LHCb and Belle II.
% may challenge our theoretical and experimental understanding
%
Lattice simulations play a central role in controlling theoretical uncertainties
due to non-perturbative aspects of QCD~\cite{Lat18:Hashimoto}.
So far, only a few modern studies have been performed 
for the $B\!\to\!D$~\cite{B2D:Nf2+1:Fermilab/MILC,B2D:Nf2+1:HPQCD}
and 
$B\!\to\!D^*$ decays~\cite{B2Dstar:Nf2+1:Fermilab/MILC:w1,B2Dstar:Fermilab/MILC:lat17,B2Dstar:HPQCD:w1,B2Dstar:Fermilab/MILC:lat18}
on gauge ensembles with staggered-type sea quarks. 
We also note that 
only preliminary results are available for 
$B\!\to\!D^*$ at nonzero recoil~\cite{B2Dstar:Fermilab/MILC:lat17,B2Dstar:Fermilab/MILC:lat18}.

The JLQCD Collaboration is pursuing a series of studies of
$B$ meson decays~\cite{Lat15:Fahy,B2Kll:JLQCD:lat18,B2pi:JLQCD:lat18,incl:JLQCD:lat18}
including the $B\!\to\!\pi\ell\nu$~\cite{B2pi:JLQCD:lat18}
and inclusive decays~\cite{incl:JLQCD:lat18},
which are relevant to the tension in the CKM matrix elements.
In this article,
we report on our on-going calculation of
the $B\!\to\!D^{(*)}$ form factors at zero and non-zero recoils.

%// simulation =================================================================

\section{Simulation}
\label{sec:sim}

We simulate 2+1 flavor QCD
using the tree-level improved Symanzik gauge action
and the M\"obius domain-wall quark action~\cite{MDWF}.
A careful choice of the detailed structure of the latter~\cite{MDWF:JLQCD:lat13}
enables us to preserve chiral symmetry to good accuracy
at moderately large lattice cutoff $a^{-1}\!\simeq\!2.5$\,--\,4.5~GeV.
This simplifies the renormalization of the relevant weak currents.
We simulate the strange quark mass $m_s$ close to its physical value,
whereas the degenerate up and down quark mass $m_{ud}$
corresponds to pion masses as low as $M_\pi\!\sim\!310$~MeV.
In this article,
we present our results at three combinations of
$(a^{-1},m_{ud},m_s)$ listed in Table~\ref{tbl:sim:param}.
At each $(a^{-1},m_{ud},m_s)$,
the spatial lattice size $L$ satisfies a condition $M_\pi L \!\gtrsim\!4$ 
to control finite volume effects,
and the statistics are 5,000 Molecular Dynamics time.
We note that
calculations at a larger cutoff $a^{-1}\!\sim\!4.5$~GeV
and a lighter $M_\pi\!\sim\!230$~MeV are underway. 

At the moderately large cutoffs $a^{-1}\!\geq\!2.5$~GeV,
we employ the same action for charm and bottom quarks.
The charm quark mass $m_c$ is set to its physical value,
whereas we take bottom quark masses
$m_b\!=\!1.25^2m_c$ and $1.25^4m_c$ if $m_b\!\leq\!0.8\,a^{-1}$. 
From our studies of the $B$ and $D$ meson (semi)leptonic decays~\cite{Lat15:Fahy,B2pi:JLQCD:lat18,D2Kpi:JLQCD:lat17},
discretization errors are not expected to be large with this setup.

\begin{table}[b]
\centering
\small
% \begin{flushleft}
\caption{
  Simulation parameters.
  $N_s^3\!\times\!N_t\!\times\!N_5$ represents the five dimensional lattice size
  for the domain-wall formulation.
  Quark masses are bare value in lattice units.
}
% \end{flushleft}
\vspace{0mm}
\label{tbl:sim:param}
\begin{tabular}{lllllllll}
   \hline 
   $\beta$ & $N_s^3\!\times\!N_t\!\times\!N_5$ & $a^{-1}$[GeV] 
           & $m_{ud}$ & $m_s$ & $M_\pi$[MeV] & $M_K$[MeV] & $m_b/m_c$
           & $\dt+\dtp$ 
   \\ \hline
   4.17  & $32^3\!\times\!64\!\times\!12$ & 2.453(4) 
         & 0.019 & 0.040 & 499(1) & 618(1) & $1.25^2$
         & 24, 28
   \\
   4.17  & $32^3\!\times\!64\!\times\!12$ & 2.453(4) 
         & 0.007 & 0.040 & 309(1) & 547(1) & $1.25^2$
         & 22, 26
   \\ \hline
   4.35  & $48^3\!\times\!96\!\times\!8$  & 3.610(9)
         & 0.012 & 0.025 & 501(2) & 620(2) & $1.25^2$, $1.25^4$
         & 36, 42
   \\ \hline
\end{tabular}
% \vspace{0mm}
\end{table}

%// form factors ===============================================================

\section{Form factors}

%// FFs

The $B\!\to\!D$ decay proceeds only through the weak vector current $V_\mu$
due to parity symmetry of QCD, 
whereas the axial current $A_\mu$ also contributes to $B\!\to\!D^*$.
The relevant matrix elements are parametrized by six form factors in total:
\bea
   \sqrt{M_B M_D}^{-1}
   \langle D(\pp) | V_\mu | B(p) \rangle
   & = &
   (v+\vp)_\mu h_+(w) + (v-\vp)_\mu h_-(w),
   \\[2mm]
   \sqrt{ M_B M_{D^*} }^{-1}
   \langle D^*(\epsilon,\pp) | V_\mu | B(p) \rangle
   & = &
   \varepsilon_{\mu\nu\rho\sigma} \, \epsilon^{*\nu} v^{\prime\rho} v^\sigma \, h_V(w),
   \\[2mm]
   \sqrt{ M_B M_{D^*} }^{-1}
   \langle D^*(\epsilon,\pp) | A_\mu | B(p) \rangle
   & = &
   -i(w+1) \, \epsilon_\mu^* \,   h_{A_1}(w)
   + i(\epsilon^* v)\, v_\mu \, h_{A_2}(w) + i(\epsilon^* v)\, \vp_\mu \, h_{A_3}(w),
   \hspace{10mm}
\eea
where $v\!=\!p/M_B$ and $\vp\!=\!\pp/M_{D^{(*)}}$ are
the four velocity of $B$ and $D^{(*)}$, 
$w=v\vp$ is the recoil parameter,
and $\epsilon$ is the polarization vector of $D^*$
satisfying $\epsilon^*\pp\!=\!0$.
In this study,
the $B$ meson is at rest ($\bfp\!=\!\bfz$),
and we change the three momentum of $D^{(*)}$ as 
$|\bfpp|^2\!=\!0,1,2,3$ (in units of $(2\pi/L)^2$)
to study the $w$ dependence of the form factors.

%// 3- and 2-pt functions

These matrix elements can be extracted from the asymptotic behavior
of three-point functions
\bea
   &&
   C_{\mathcal O_\Gamma}^{BD^{(*)}}(\dt,\dtp;\bfp,\bfpp)
   \nn \\
   & = & 
   \frac{1}{N_{\tsrc}}\sum_{\tsrc}
   \sum_{\bfxsrc,\bfx,\bfxp}
   \langle 
      {\mathcal O}_{D^{(*)}}(\bfxp,\tsrc+\dt+\dtp)
      {\mathcal O}_\Gamma(\bfx,\tsrc+\dt)
      {\mathcal O}_{B}(\bfxsrc,\tsrc)^\dagger
   \rangle
   e^{-i\bfp(\bfx-\bfxsrc)-i\bfpp(\bfxp-\bfx)}
   \nn \\
   & \to &
   % & \xrightarrow[\dt,\dtp \to \infty ]{} &
   \frac{Z_{D^{(*)}}^*(\bfpp)\,Z_B(\bfp)}{4E_{D^{(*)}}(\bfpp)E_B(\bfp)}
   \langle D^{(*)}(\pp) | {\mathcal O}_\Gamma | B(p) \rangle
   e^{-E_{D^{(*)}}(\bfpp)\dtp -E_B(\bfp)\dt }
   \hspace{3mm} (\dt, \dtp \to \infty),
   \label{eqn:ff:corr_3pt}
\eea
where ${\mathcal O}_\Gamma\!=\!V_\mu$ or $A_\mu$,
and the argument $\epsilon$ is suppressed for $Z_{D^*}$
and $|D^*(\pp)\rangle$.
Gaussian smearing is applied to the interpolating field 
${\mathcal O}_P$ ($P\!=\!B,D,D^*$) to enhance its overlap
to the ground state
$Z_P(\bfp)\!=\!\langle P(p) | {\mathcal O}_P^\dagger \rangle$.
We take two values of the total temporal separation $\dt+\dtp$
listed in Table~\ref{tbl:sim:param}
to check whether the excited contamination is sufficiently suppressed.
The two-point function
\bea
   C^{P}(\dt;\bfp)
   & = & 
   \frac{1}{N_{\tsrc}}\sum_{\tsrc}
   \sum_{\bfxsrc,\bfx}
   \langle 
      {\mathcal O}_{P}(\bfx,\tsrc+\dt)
      {\mathcal O}_{P}(\bfxsrc,\tsrc)^\dagger
   \rangle
   e^{-i\bfp(\bfx-\bfxsrc)}
   \to
   \frac{|Z_P(\bfp)|^2}{2E_P(\bfp)} e^{-E_P(\bfp)\dt}
   \hspace{3mm}
   \label{eqn:ff:corr_2pt}
\eea
is also measured to estimate
the rest mass $M_P$, energy $E_P$ and the overlap factor $Z_P$
($P\!=\!B,D,D^*$).

%// improvement of statistical accruacy: average

We improve the statistical accuracy of the three- and two-point functions,
$C_{{\mathcal O}_\Gamma}^{BD^{(*)}}$ and $C^P$,
by averaging over the source location $(\bfxsrc,\tsrc)$.
For the temporal location $\tsrc$, 
we simply repeat our calculation at two different time-slices
(hence, $N_\tsrc\!=\!2$
in Eqs.~(\ref{eqn:ff:corr_3pt}) and (\ref{eqn:ff:corr_2pt})).
The summation over the spatial location $\bfxsrc$
is implemented by using the volume source with $Z_2$ noise.
We also average $C_{{\mathcal O}_\Gamma}^{BD^{(*)}}$ and $C^P$
over the momentum configurations,
which are equivalent due to rotational and parity symmetries.
These procedures improve the statistical accuracy
by factor of 2\,--\,4 with our simulation setup.

%// improvement of statistical accruacy: ratios

We construct ratios of $C_{{\mathcal O}_\Gamma}^{BD^{(*)}}$ and $C^P$
for a more precise and reliable calculation of the form factors.
The double ratios without nonzero momentum~\cite{double_ratio}
\bea
   R_{1(i)}^{BD^{(*)}}(\dt,\dtp)
   & = &
   \frac{ C_{V_4(A_i)}^{BD^{(*)}}(\dt,\dtp;\bfz,\bfz)\,
          C_{V_4(A_i)}^{D^{(*)}B}(\dt,\dtp;\bfz,\bfz) }
        { C_{V_4(A_i)}^{BB}(\dt,\dtp;\bfz,\bfz)\,
          C_{V_4(A_i)}^{DD}(\dt,\dtp;\bfz,\bfz) }
   \xrightarrow[\dt,\dtp\to\infty]{}
   |h_{+(A_1)}(1)|^2 
   \label{eqn:ff:b2d+b2d*:r1}
\eea
give an accurate estimate of $h_+$ and $h_{A_1}$
at zero recoil $w\!=\!1$,
which are important inputs in the conventional determination of $|V_{cb}|$.
The analysis of $h_+$ and $h_-$ at nonzero recoil $w\!>\!1$
is analogous to our study of $K\!\to\!\pi$ and
previous studies of $B\!\to\!D$~\cite{K2pi:JLQCD,B2D:Nf2+1:Fermilab/MILC,B2D:Nf2+1:HPQCD}.
Together with ratios
\bea
   R_2^{BD}(\dt,\dtp;\bfz,\bfpp)
   & = &
   \frac{ C_{V_4}^{BD}(\dt,\dtp;\bfz,\bfpp)\, C^{D}(\dtp,\bfz)  }
        { C_{V_4}^{BD}(\dt,\dtp;\bfz,\bfz)\,  C^{D}(\dtp,\bfpp) }
   \to
   \frac{(1+w)h_+(w) + (1-w)h_-(w)}{2h_+(1)},
   \label{eqn:ff:b2d:r2}
   \\ 
   R_{3i}^{BD}(\dt,\dtp;\bfz,\bfpp)
   & = &
   \frac{ C_{V_i}^{BD}(\dt,\dtp;\bfz,\bfpp) }
        { C_{V_4}^{BD}(\dt,\dtp;\bfz,\bfpp) }
   \to
   v^\prime_i
   \frac{ h_+(w) - h_-(w) }{ (1+w)h_+(w) + (1-w)h_-(w) },
   \label{eqn:ff:b2d:r3}
\eea
we can reconstruct the form factors as
\bea
   h_{+(-)}(w)
   & = &
   \sqrt{R_1^{BD}} R_2^{BD}
   \left\{ 1 \pm (1 \mp w) \frac{R_{3i}^{BD}}{\vp_i} \right\}.
   \label{eqn:ff:b2d:h+-}
\eea

%// ratios for B->D*

The analysis of $B\!\to\!D^*$ is slightly more involved,
and we need to distinguish the $D^*$ momentum $\bfppnp$,
which induces $\epsilon^*\!v\!\ne\!0$ through the convention
$\epsilon^*\!\pp\!=\!0$,
and $\bfppp$ leading to $\epsilon^* v\!=\!0$ 
(note that ${\bf v}\!=\!\bfz$ in this study).
The $w$-dependence of $h_{A_1}$ is studied
by a ratio similar to (\ref{eqn:ff:b2d:r2}) with $A_i$ and $\bfppp$
\bea
   R_{2i}^{BD^*}(\dt,\dtp;\bfz,\bfppp)
   & = &
   \frac{ C_{A_i}^{BD^*}(\dt,\dtp;\bfz,\bfppp)\,C^{D^*}(\dt,\bfz)  }
        { C_{A_i}^{BD^*}(\dt,\dtp;\bfz,\bfz)\,C^{D^*}(\dt,\bfppp) }
   \to
   \frac{1+w}{2} \frac{h_{A_1}(w)}{h_{A_1}(1)},
   \label{eqn:ff:b2d*:r2}
\eea   
whereas
a ratio $C_{A_{\{i,4\}}}^{BD^*}(\dt,\dtp;\bfz,\bfppnp)/
         C_{A_i}^{BD^*}(\dt,\dtp;\bfz,\bfppp)$
is sensitive to $h_{A_2}$ and $h_{A_3}$ at $w\!>\!1$~\cite{B2Dstar:Fermilab/MILC:lat17}.
A form factor ratio $R_1(w)\!=\!h_V(w)/h_{A_1}(w)$
is a key quantity in recent phenomenological discussions
about the tension in $|V_{cb}|$,
and is determined from
\bea
   R_{3i}^{BD^*}(\dt,\dtp;\bfz,\bfppp)
   & = &
   \frac{ C_{V_i}^{BD^*}(\dt,\dtp;\bfz,\bfppp) }
        { C_{A_i}^{BD^*}(\dt,\dtp;\bfz,\bfppp) }
   \to
   \frac{ \epsilon_{ijk} \epsilon^*_j v_{\perp k}^\prime}{1+w}
   \frac{h_V(w)}{h_A(w)}.
   \label{eqn:ff:b2d*:r3}
\eea
We average $C_{V_i}^{BD^{(*)}}$ and $C_{A_i}^{BD^*}$ 
over $i\!=\!1,2,3$ with appropriately rotated $\bfpp$ and $\bfppnp$
before calculating the above ratios.
Note that
renormalization factors cancel even in the ratio~(\ref{eqn:ff:b2d*:r3})
due to chiral symmetry preserved in our simulations.

%// ratio example

\begin{figure}[t]
\begin{center}
   \includegraphics[angle=0,width=0.48\linewidth,clip]%
                   {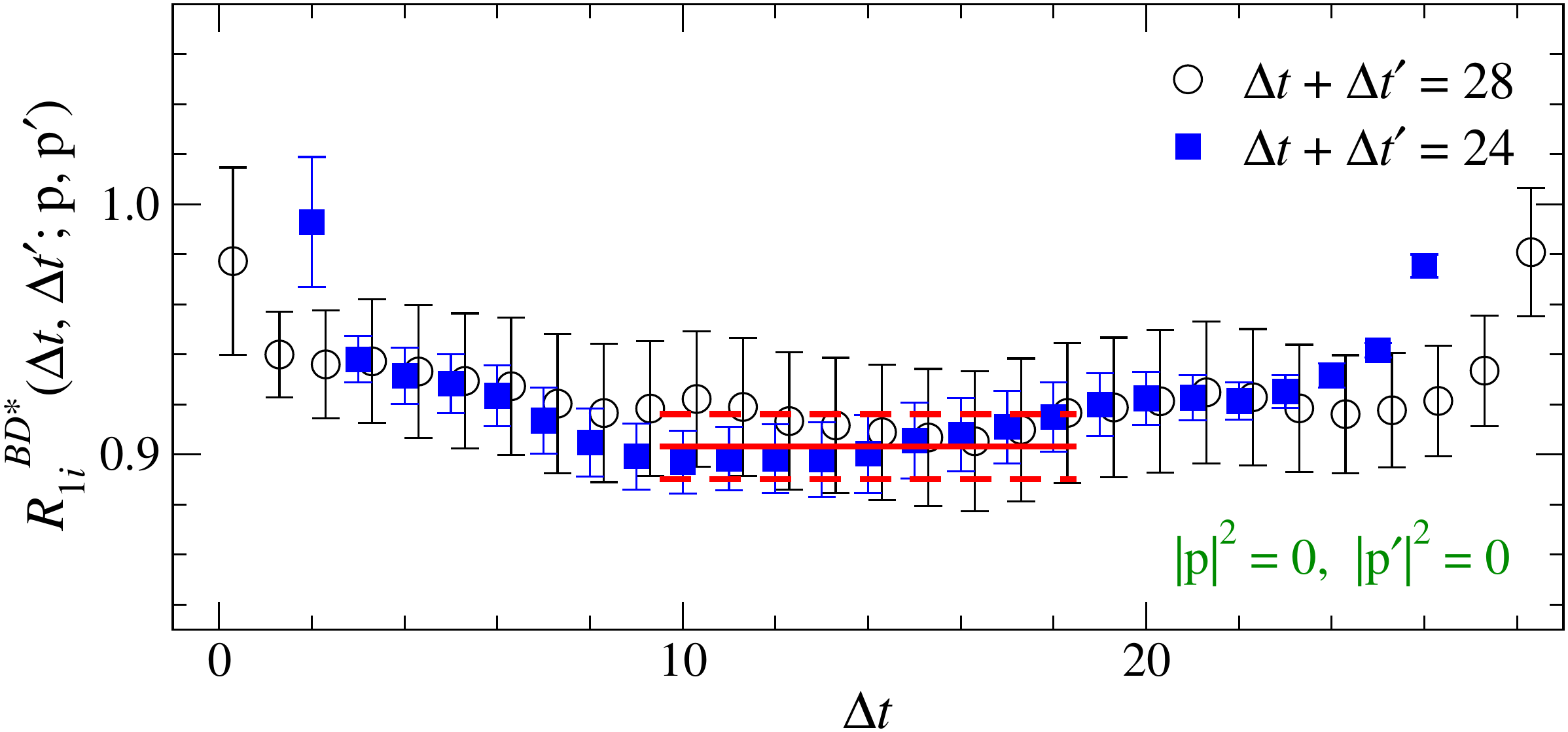}
   \hspace{3mm}
   \includegraphics[angle=0,width=0.48\linewidth,clip]%
                   {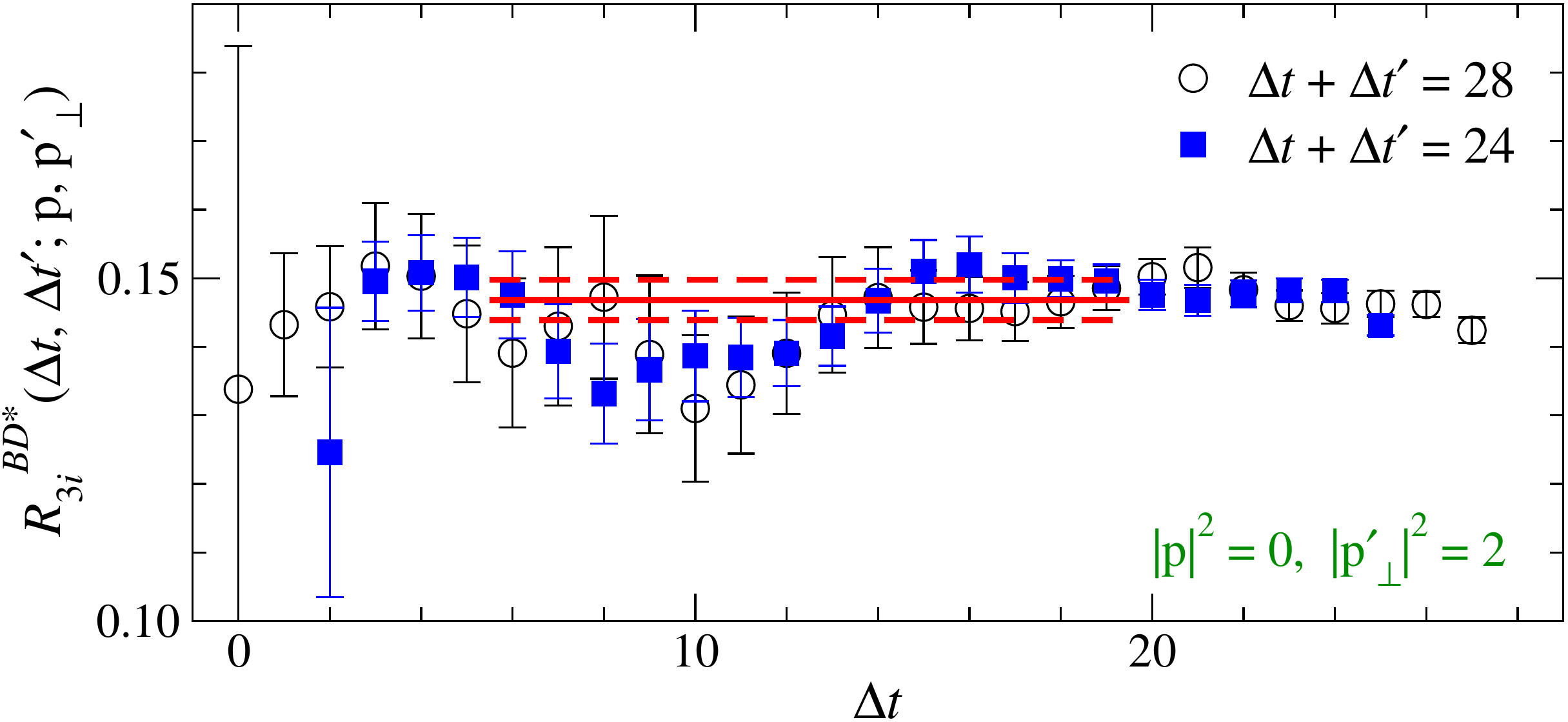}

   \vspace{-2mm}
   \caption{
      Ratios of three-point functions for $B\!\to\!D^*$.
      The left and right panels show
      $R_{1i}^{BD^*}(\dt,\dtp;\bfz,\bfz)$ and
      $R_{3i}^{BD^*}(\dt,\dtp;\bfz,\bfppp)$, respectively,
      as a function of $\dt$.
      We plot data at $\beta\!=\!4.17$ and $(m_{ud},m_s)\!=\!(0.019,0.040)$.
      The open circles and filled squares show data
      with $\dt+\dtp\!=\!28$ and 24, respectively.
      The red horizontal lines show a constant fit
      to data with $\dt+\dtp\!=\!24$.
   }
   \label{fig:ff:rat}
\end{center}
\vspace{0mm}
\end{figure}

Figure~\ref{fig:ff:rat} shows an example of the ratios
for $B\!\to\!D^*$.
We confirm reasonable consistency between 
two sets of data with different values of $\dt+\dtp$
suggesting that the excited state contamination is sufficiently suppressed.
The statistical accuracy with the smaller value of $\dt+\dtp\!\approx\!1.8$~fm
is typically 2\,--\,6\,\%
for $h_+$, $h_{A_1}$, $h_{A_3}$, $h_V$,
which are reduced to the Isgur-Wise function $\xi(w)$
with the normalization $\xi(1)\!=\!1$
in the heavy quark limit $m_c, m_b \!\to\!\infty$.
Other form factors $h_-$ and $h_{A_2}$ vanish in the heavy quark limit,
and their results are close to zero
with a typical accuracy of $\lesssim\!50$\,\%.

%// FF vs w 

\begin{figure}[t]
\begin{flushleft}
   \includegraphics[angle=0,width=0.48\linewidth,clip]%
                   {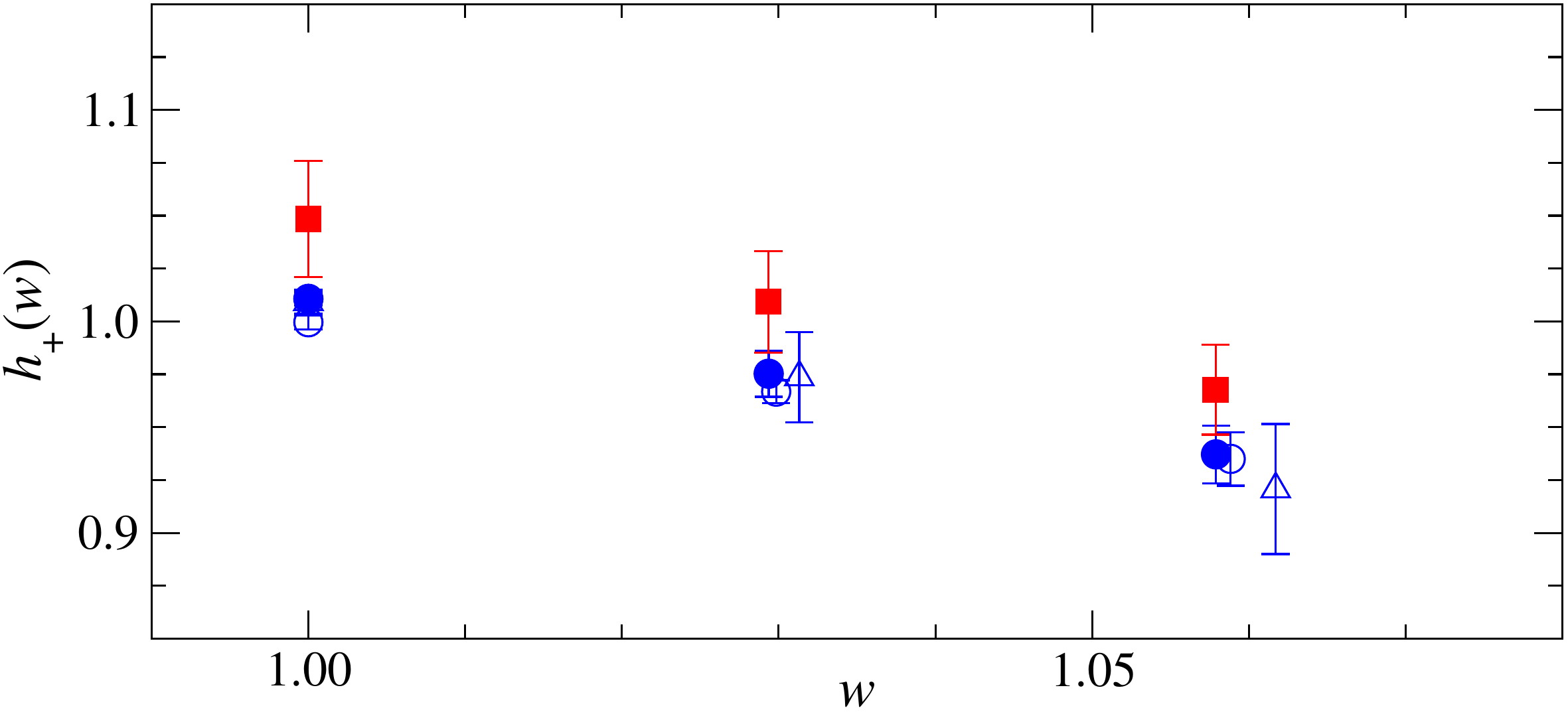}
   \hspace{3mm}
   \includegraphics[angle=0,width=0.48\linewidth,clip]%
                   {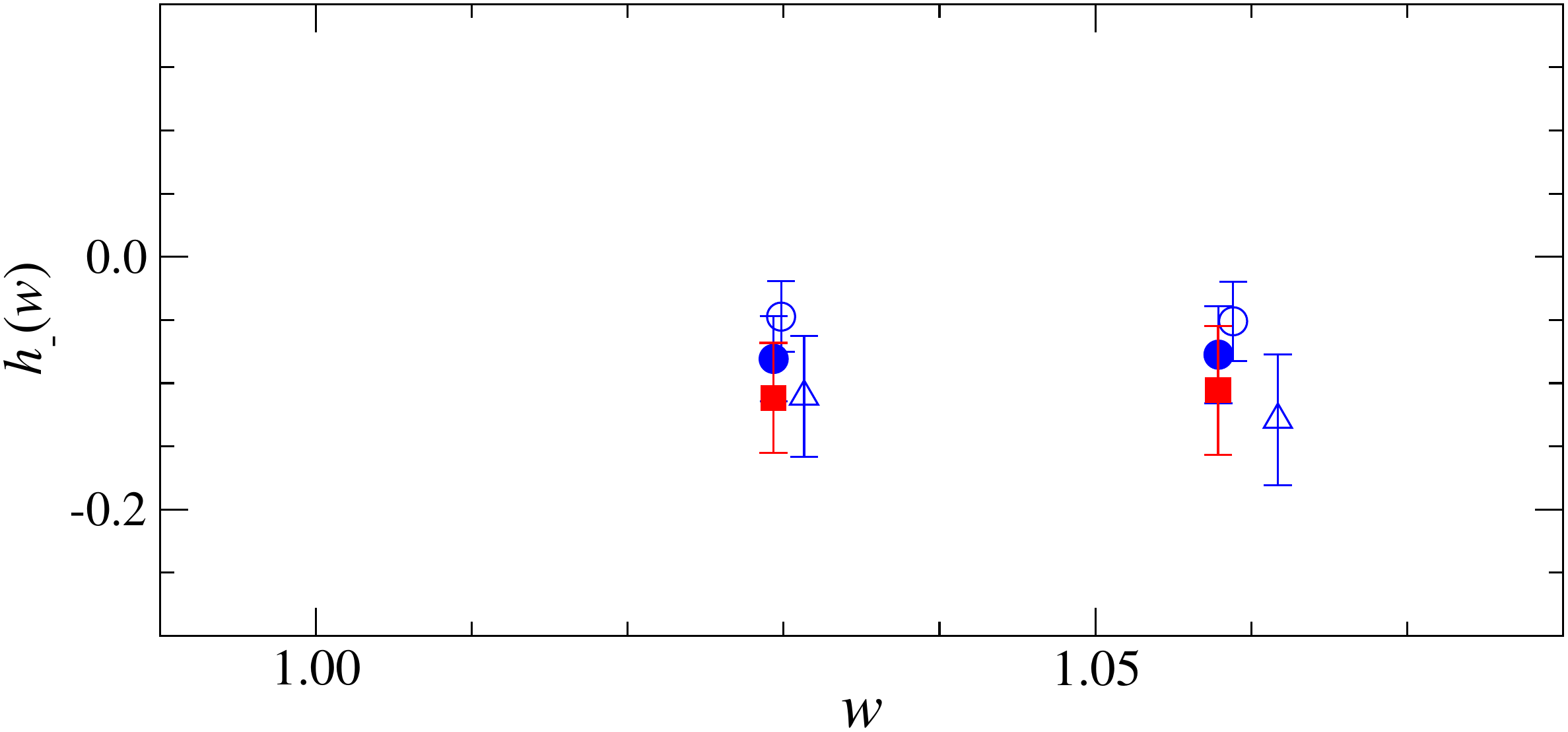}
   \vspace{1mm}

   \includegraphics[angle=0,width=0.48\linewidth,clip]%
                   {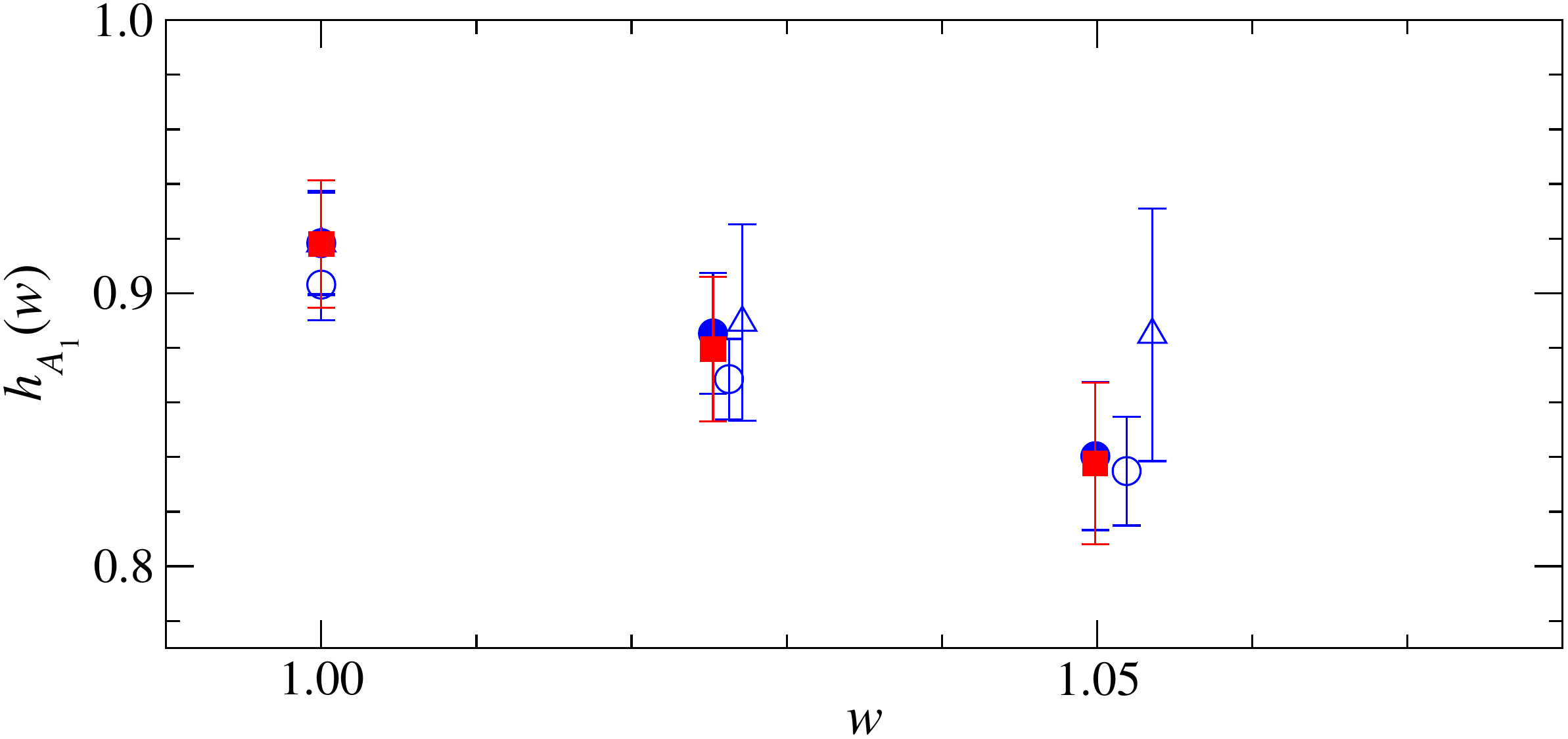}
   \hspace{3mm}
   \includegraphics[angle=0,width=0.48\linewidth,clip]%
                   {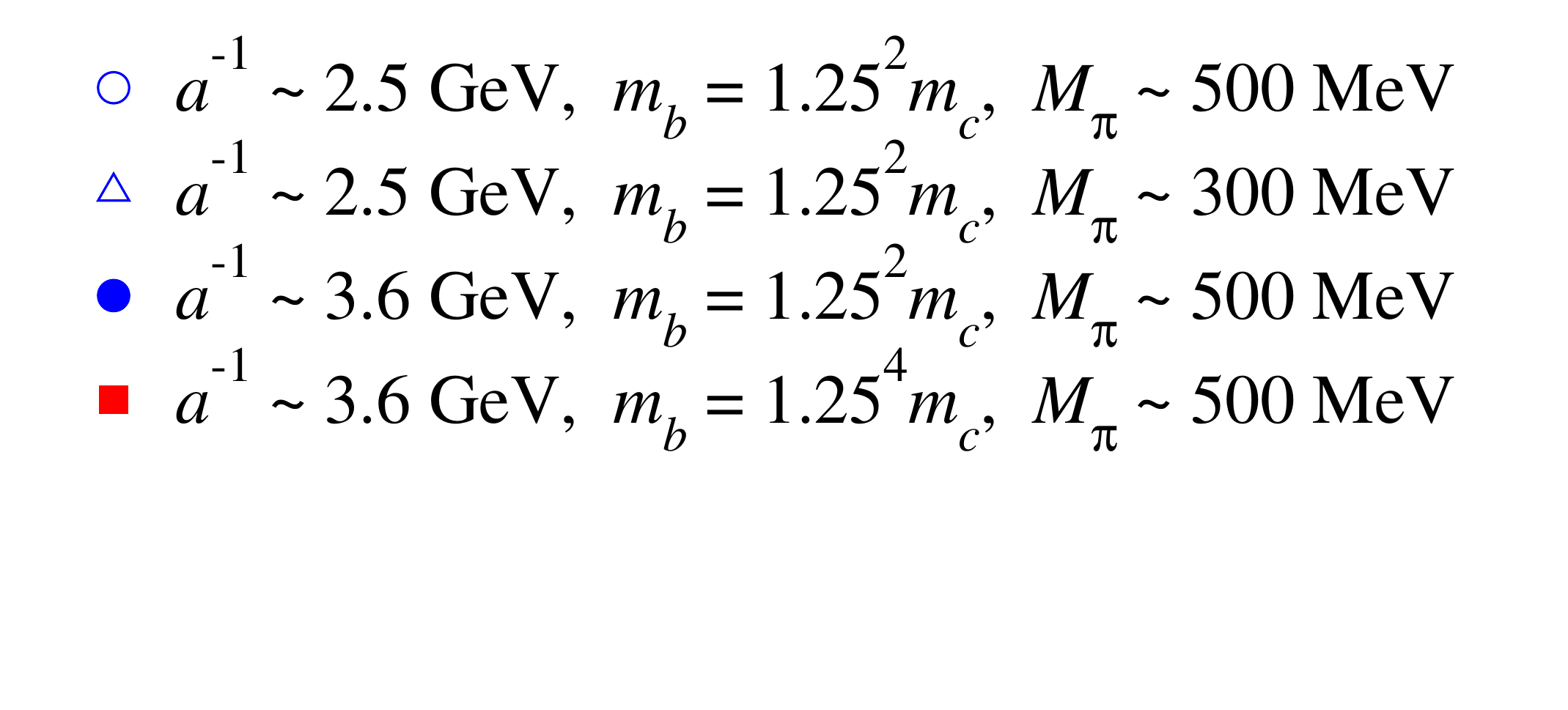}

   \vspace{-2mm}
   \caption{
      Form factors as a function of $w$.
      We plot $h_+$, $h_-$ and $h_{A_1}$
      in the top-left, top-right and bottom-left panels, respectively.
      The blue (red) symbols show data with $m_b\!=\!1.25^2 m_c$ ($1.25^4 m_c$),
      whereas the open (filled) symbols are at $a^{-1}\!\simeq\!2.5$ (3.6) GeV.
   }
   \label{fig:ff:w-dep}
\end{flushleft}
\vspace{0mm}
\end{figure}

Figure~\ref{fig:ff:w-dep} shows 
results for $h_+$, $h_-$ and $h_{A_1}$ at different simulation points
as a function of $w$.
These form factors describe the differential decay rates at zero recoil
$d\Gamma/dw(B\!\to\!D^{(*)}\ell\nu)|_{w=1}$
for the massless lepton $m_\ell\!=\!0$.
These and other form factors mildly depend
on $a^{-1}$, $m_b$ and $M_\pi$
-- at least in our simulation range of these parameters.
We note that similar mild dependence on $a^{-1}$ and $m_b$ is also observed
for the $B\!\to\!\pi\ell\nu$ form factors~\cite{B2pi:JLQCD:lat18}.
While all the form factors have to be extrapolated to
the continuum limit and physical up, down and bottom quark masses,
the mild dependence may suggest that 
the preliminary results are not far from these limits
and the extrapolation can be reasonably controlled.

%// heavy quark symmetry violation =============================================

\section{Heavy quark symmetry violation and $|V_{cb}|$}

%// history 

The $B\!\to\!D^*\ell\nu$ differential decay rate for $m_\ell\!=\!0$
is described by three combinations of the form factors,
$h_{A_1}$, $h_V$ and $r h_{A_2}+ h_{A_3}$ ($r\!=\!M_{D^*}/M_B$).
Boyd, Grinstein and Lebed (BGL) proposed a model independent
parametrization~\cite{B2D*:FF:BGL},
which Taylor-expands the (regularized) form factors
around zero recoil in $w-1$,
or in terms of a small kinematical parameter
$z\!=\!(\sqrt{w+1}-\sqrt{2}a)/(\sqrt{w+1}+\sqrt{2}a)$
with $a$ a tunable input.
While one can derive constraints on the expansion parameters from unitarity,
they are rather weak.
The conventional determination of $|V_{cb}|$ therefore employs
the Caprini-Lellouch-Neubert (CLN) parametrization~\cite{B2D*:FF:CLN},
which has only four free parameters: 
the normalization and slope of $h_{A_1}$,
$R_1(1)\!=\!h_V(1)/h_{A_1}(1)$ and
$R_2(1)\!=\!\left\{r h_{A_2}(1)+ h_{A_3}(1)\right\}/h_{A_1}(1)$.
The remaining parameters are constrained by
% heavy quark symmetry
next-to-leading order (NLO) heavy quark effective theory (HQET)
with QCD sum rule inputs for the sub-leading Isgur-Wise functions.
Recently, Belle has published 
a preliminary analysis of the differential decay rate
with unfolded kinematical and angular distributions
for the first time~\cite{B2D*:exprt:Belle:unfolded}\footnote{
Reference~\cite{B2D*:exprt:Belle:unfolded} analyzes
results with a tagged approach.
We note that, after the conference,
Belle updated their analysis of $B\!\to\!D^*\ell\nu$
with an untagged approach  
by using both the BGL and CLN parametrizations~\cite{B2D*:exprt:Belle:unfolded2}.
}.
This allows a determination based on the BGL parametrization
yielding $|V_{cb}|\!\times\!10^3\!=41.7(^{+2.0}_{-2.1})$~\cite{Vcb:BGS}
and $41.9(^{+2.0}_{-1.9})$~\cite{Vcb:GK},
which are compatible with the inclusive determination 42.0(0.5)
and slightly larger than 38.2(1.5)
with the CLN parametrization~\cite{B2D*:exprt:Belle:unfolded}.
This led to recent phenomenological discussions
about higher order correction to NLO HQET~\cite{Vcb:BGS:2,Vcb:BLPR}.

%// HQS violation

\begin{figure}[t]
\begin{flushleft}
   \includegraphics[angle=0,width=0.47\linewidth,clip]%
                   {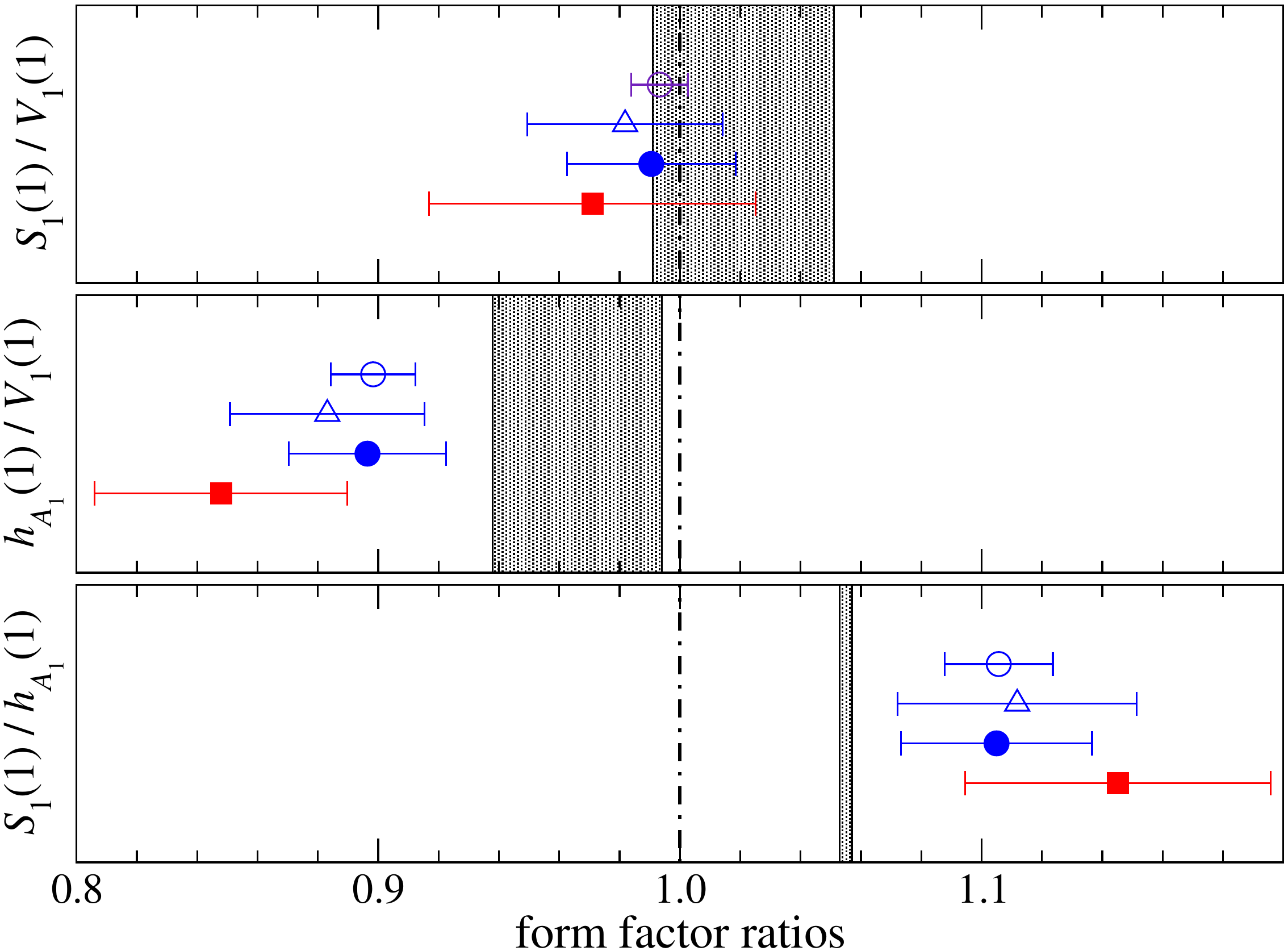}
   \hspace{3mm}
   \includegraphics[angle=0,width=0.49\linewidth,clip]%
                   {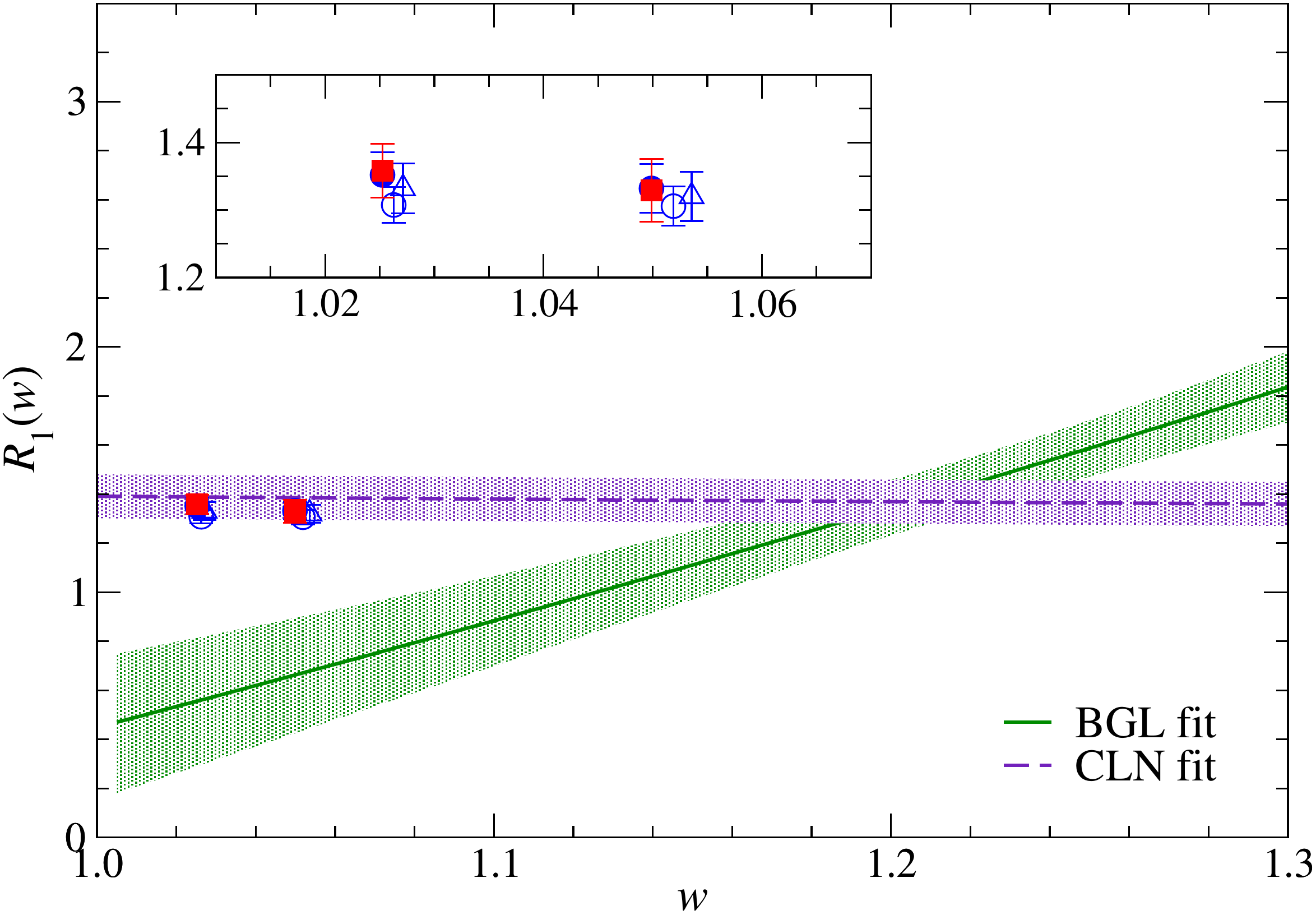}

   \vspace{-2mm}
   \caption{
      Left panel: form factor ratios
      $S_1(1)/V_1(1)$ (left-top panel),
      $h_{A_1}(1)/V_1(1)$ (left-middle panel) and 
      $S_1(1)/h_{A_1}(1)$ (left-bottom panel).
      These ratios are unity in the heavy quark limit,
      and the shaded region shows the NLO HQET estimate~\cite{RD*:CLN:BLPR}.
      Right panel:
      $R_1(w)$ as a function of $w$.
      The BGL and CLN fits shown in the green and purple bands, respectively,
      are from Ref.~\cite{Vcb:BLPR} by courtesy of the authors.
      The inner panel magnifies a small region around our lattice results.
      In all the panels, 
      our results are plotted by 
      the same symbols as Fig.~\protect\ref{fig:ff:w-dep}.
   }
   \label{fig:vcb:hqs+cln}
\end{flushleft}
\vspace{0mm}
\end{figure}

In the left panel of Fig.~\ref{fig:vcb:hqs+cln},
we compare form factor ratios between lattice QCD and NLO HQET.
We again confirm that our lattice results mildly depend on
$a^{-1}$, $m_b$ and $M_\pi$.
There seems to be a systematic deviation for $h_{A_1}/V_1(1)$
and $S_1(1)/h_{A_1}(1)$,
where
$V_1\!=\!h_+-(1-r)\,h_-/(1+r)$ and 
$S_1\!=\!h_+-(1+r)(w-1)\,h_-/(1-r)(w+1)$
are vector and scalar form factors for $B\!\to\!D\ell\nu$.
Note that
the CLN constraint on $h_{A_1}$ is derived
from $h_{A_1}(w)/V_1(w)$ in NLO HQET and the unitarity bound for $V_1(w)$
for $B\!\to\!D\ell\nu$~\cite{B2D*:FF:CLN}.
Our observation suggests that NLO HQET may receive 
significant higher order corrections as discussed in Ref.~\cite{Vcb:BGS:2}.

However, this seems not to be the case for $R_1(w)$,
which exhibits one of the largest differences
between the CLN and BGL analyses~\cite{Vcb:BLPR}.
The right panel of Fig.~\ref{fig:vcb:hqs+cln} shows that
our results for $R_1(w)$ favor the CLN prediction, 
though they eventually have to be extrapolated to
the continuum limit and the physical quark masses.
These observations suggest that, at the moment,
the $|V_{cb}|$ tension may not be simply attributed
to the higher order corrections to NLO HQET,
and more lattice data are welcome for a more detailed comparison
between the CLN and BGL analyses.

%// conclusion =================================================================

\section{Outlook}

In this article,
we report on our study of the $B\!\to\!D^{(*)}\ell\nu$ decays
at zero and nonzero recoils.
With our simulation setup,
the relevant form factors show mild dependence
on $a^{-1}$, $m_b$ and $M_\pi$,
which led us to discuss implication of the preliminary results
to the $|V_{cb}|$ tension. 
Our goal is to obtain purely theoretical prediction
for the form factors through lattice simulations
and a model-independent parametrization such as BGL
towards a more reliable determination of $|V_{cb}|$.
To this end,
we are planning to extend our simulation
to a lighter pion mass $M_\pi\!\sim\!230$~MeV, 
a finer lattice with $a^{-1}\!\sim\!4.5$~GeV
and $m_b\!=\!1.25^5 m_c$
for a controlled extrapolation of our results
to the continuum limit and physical $m_{ud}$ and $m_b$.
Our data at different $m_b$'s are expected to be useful
to test the heavy quark scaling for the form factors
given by heavy quark symmetry.

%// Acknowledgment

We are grateful for F.U.~Bernlochner, Z.~Ligeti, M.~Papucci, and D.J.~Robinson
for making their numerical results in Ref.~\cite{Vcb:BLPR}
available to us.
Numerical simulations are performed on Oakforest-PACS
at JCAHPC under a support of the HPCI System Research Projects
(Project IDs: hp170106 and hp180132)
and Multidisciplinary Cooperative Research Program
in CCS, University of Tsukuba
(Project IDs: xg17i036 and xg18i016).
This work is supported in part by JSPS KAKENHI Grant Numbers
16K05320, 18H01216, 18H03710 and 18H04484.

%// references =================================================================

\end{document}